# Whitepaper

## DISPENSER CONCEPT FOR UNMANNED AERIAL VEHICLES, PLATFORMS AND SYSTEMS (UAV/UAS/Drone).

**Author:** Manan Suri

**V1:** 20th May 2016

## Abstract

System, design and methodology to load and dispense different articles from an autonomous aircraft are disclosed. In one embodiment, the design of a unique detachable dispenser for delivery of articles is described along with an intelligent methodology of loading and delivering the articles to and from the dispenser. Design of the dispenser, interaction of the dispenser with the flight control unit and ground control or base-station, and interaction of the base station with the sender or recipient of the article, are also described.

## Background

The subject matter described herein relates to delivery of one or more articles at specific targets, including but not limited to mail, postage, letters, newspapers, magazines, post cards, packages etc. from fixed wing, rotor based, or any other type of unmanned air borne vehicles or systems.

Aircraft particularly small and mid-size, autonomous or semi-autonomous, rotor- or fixed-wing based, may use the invention to drop or lower or release mail and/or packages at designated targets. There exists a strong need for smart air-borne systems to deliver mail, packages and articles to desired targets.

### *About Drone/UAV Technology*

An unmanned aerial vehicle (UAV), commonly known as a *drone* or as an unmanned aircraft/aerial system (UAS), is an aircraft without a human pilot aboard. The flight of UAVs may operate with various degrees of autonomy: either under remote control by a human operator, or fully or partially autonomously, by onboard computers/on-ground computers.

Compared to manned aircrafts, UAVs are often preferred for missions that are either too routine or too dangerous for humans. They originated mostly for military purposes, although their use is fast expanding in law-enforcement, commercial, scientific and even recreational domains. For example, applications such as policing and surveillance, high-definition (HD) aerial photography/videography, agriculture, automated delivery, and even drone racing sports. Currently civilian drones vastly outnumber military drones, with estimates of over a million sold by 2015.

The UAV global military market is dominated by pioneers United States and Israel. The US held a 60% military-market share in 2006. It operated over 9,000 UAVs in 2014. From 1985 to 2014, exported drones came predominantly from Israel (60.7%) and the United States (23.9%); top importers were The United Kingdom (33.9%) and India (13.2%) [a]. Northrop Grumman and General Atomics are the dominant manufacturers for high-end military drone/UAV systems.

The leading civil UAV companies are currently (Chinese) DJI with $500m global sales, (French) Parrot with $110m and (US) 3DRobotics with $21.6m in 2014 [b]. As of February 2016, about 325,000 civilian drones were registered with the U.S. FAA, though it is estimated more than a million have been sold in the United States alone [c].

### *Delivery Application*

UAVs have been proposed for delivery applications. For example, medicines and medical specimens into and out of inaccessible regions. In 2013, in a research project of DHL, a small quantity of medicine was delivered via a UAV.

Initial attempts at commercial use of UAVs, such as the Tacocopter company for food delivery, were blocked by FAA regulation. A 2013 announcement that online ecommerce giant Amazon was planning deliveries using UAVs was met with skepticism. In 2014, the government of UAE announced that they plan to launch a fleet of UAVs to deliver official documents and supply emergency services at accidents. Technology giants like Google, small startups like Matternet and Firefly have also entered the drone delivery segment although they are still in a nascent R & D phase. Government postal departments of different countries such as Singapore, Australia and Germany have all run small pilot projects on drone delivery.

### *Our Novelty*

By far, most of the above mentioned examples involve some proof-of-concept deliveries using UAVs for single packages and a single pre-determined address. Such an approach, single-package-to-single-address, while technically easy to demonstrate, is grossly limiting to the overall scope of the technology. The full commercial, environmental and societal benefits and potential of UAV based automated delivery can only be realized when a single UAV is designed, engineered, optimized and operated in such a way that it can be used for multiple intelligent deliveries. A revolutionary landmark will be reached, when a delivery drone can drop multiple articles like a postman. In this patent document, we propose an invention for societal, commercial and environmental gains: a *first of its kind, multi-article UAV delivery system* and its associated operation methodology.

## Objective

The objective of the system is to benefit society at large, in several ways.

Firstly, the system employs clean electric energy, instead of fuel, burnt by the alternatives (such as trucks etc.) that it intends to replace. This is highly beneficial for mankind as well as for the environment and a key consideration in this age of global warming.

Secondly, the system intends to reduce the average delivery time for a plethora of articles from several days to just 24 hours or less. Not only does this make delivery of goods more efficient and convenient for consumers, it also has far reaching implications for critical and time-sensitive items such as medicines or blood supplies from hospitals.

Thirdly, the system intends to reduce reliance on ground-based modes of transportation (cars, trucks, bicycles etc.), thereby reducing traffic in the cities in which it operates, and making navigation for everyone, a more pleasurable pursuit.

Fourthly, the system intends to reduce sound pollution (usually caused by motor vehicles) by utilizing a much quieter rotor based technology for navigation.

Last but not the least, the system intends to empower consumers, by introducing more direct interaction (whether with senders, or receivers of articles) and by providing real-time communication protocols and tracking. In this way, the system replaces the complexity of current ground-based transportation organizations, with a cleaner and more efficient interface which provides far more accurate real-time information regarding the location and status of the article, out for delivery.

## Detailed description

*The Dispenser*

Fig. 1a shows a three dimensional view of our proposed intelligent dispenser. The dispenser is essentially a box type of structure with one or more symmetric or non-symmetric internal compartments. The number of compartments, and the dimensions of individual compartments, may be defined as per user requirement. In a modular approach, the walls of some of the individual compartments may also be removed (or added back) in order to make room for larger articles wherever necessary. Each compartment will be used to hold one or more articles such as mail, letter, postcard, newspaper, book, package, magazine etc. which is to be delivered to any specific target. Each 'compartment' may have all sides in the form of openable doors and walls, or may have some sides that are hollow i.e. open to air. In the latter case, the articles may be held inside the compartment by mechanical, magnetic, electronic or any other method. The shape of compartments shown in Fig.1a are cuboidal for reference purpose, however it may change to trapezoidal, elliptical, circular or any other polygon configuration. The shapes and dimensions of the individual compartments will vary correspondingly.

Fig 1b shows the bottom or top face of the proposed dispenser. The shaded portions are intelligently controlled orifices that will be used to load and dispense the articles in the dispenser compartments. Each compartment will have one or more orifice for loading, and or for delivering the article(s). The opening and closing (if any) of the orifices may be controlled by mechanical, magnetic, electronic, electro-mechanical, electro-magnetic or a combination of more than one of the aforementioned mechanisms. In one variant, the orifices may be hollow, leaving the compartments open to air.

In another variant, there may be an enclosure which is placed under and encompasses (one or more than one) compartment(s), and may extend across the entire length or breadth of the dispenser. Once the dispenser is ready for a drop, such an enclosure may open first at a pre-determined angle, and once the desired article is released from its compartment, it may potentially land on, and move (or slide) down the path of this enclosure (voluntarily or involuntarily) into a desired slit or opening, such as (but not limited to) a mailbox. Such an enclosure may also have sensor(s) on it, that enable it to establish a contact with the point of landing Fig 1d shows one possible iteration of such an enclosure. The dimensions of the enclosure can be chosen according to the user's specifications.

### *On-board Electronics*

Fig. 1c shows a schematic of different functional blocks of the proposed intelligent dispenser.

The heart of the system is an electronic microprocessor(s)/microcontroller (s), which controls the working of the dispenser. The processor(s) controls and interfaces with all other blocks of the dispenser such as – I/O interface, Memory, Battery/Power block, Sensors, Communication block and delivery mechanism.

The I/O interface enables communication, data and instruction exchange between the dispenser and any external system such as the aircraft flight control board, external computer or digital electronic system. The dispense and load mechanism block consists of the mechanical, magnetic, electronic, electro-mechanical, electro-magnetic apparatus to enable opening and closing (if any) of the orifices of the dispenser. It may consist of actuators, switches, ICs, MEMS circuits, gears, pulleys, motors, drives, magnets, hinges, shutters, and any other component as per requirement. If and wherever applicable, the processor issues orifice open/close instruction(s) to the delivery mechanism block.

The communication block consists of one or more systems including, but not limited to GPS, GPRS, GSM, WiFi, bluetooth, single/multi-frequency transmitter/receiver, transponder, antennas, etc. for enabling communication, transmission and receipt of data/instructions between the dispenser and any external digital or analog electronic system or platform with relevant receiver capabilities. For example, these may include and are not limited to stationary or moving communication base stations, smart devices such as smart phones, tablets, servers, PCs, other similar intelligent dispensers, satellites, cell phone towers, antennas etc.

The sensor block consists of one or a combination of multiple mechanical, electronic, electro-mechanical, acoustic, electro-magnetic, chemical sensors such as- visual, IR, temperature, humidity, thermal, multi-spectral, acoustic, cameras, pressure, weight, accelerometers, gyroscopes, etc. to sense various parameters of the dispenser, the articles inside it and its surroundings when it is stationary or in motion. In one variant, the sensor block may also comprise a sensor for detecting contraband or explosives to avoid loading of such articles onto the dispenser.

The memory block is to provide on-board storage for any data or instructions and in particular coordinates of the addresses where specific articles stored in specific dispenser compartments need to be dropped or lowered or released or unloaded. In one variant where a camera or visual recorder is attached to the dispenser, the memory block may also store previously recorded images of drop sites for enhanced accuracy in subsequent drops. The memory block will be read and programmed with the help of the microprocessor.

The battery and power management block consists of the battery and required electronic circuits used to power any sub block of the dispenser.

Fig 2 shows a representative sample configuration in which the intelligent dispenser is attached to an aircraft module or block (s). The aircraft module basically enables air-borne movement of the dispenser. The aircraft block maybe remotely piloted, autonomous or semi-autonomous. It may be based on rotary, fixed-wing or hybrid flight, landing and take-off mechanisms. For example, it may be in a standard quadcoptor, hexacoptor, octacoptor or any other configuration as per the requirements of aerodynamic stability of the dispenser and itself when in flight.

The dispenser may be attached and detached from the aircraft module mechanically, magnetically or through any other appropriate mechanism. For a rotary configuration, the aircraft module will consist of motor(s), motor controller(s), propeller(s), flight control system, sensor(s), batteries, and related electronic/mechanical components.

### Aircraft System

Fig 3 shows a schematic of the aircraft and dispenser blocks interacting with each other.

The aircraft system and dispenser will communicate with each other (if and when required) directly using wired connections and standard communication protocols such as USB etc through their respective I/O interfaces or wirelessly using their communication sub-blocks (wireless communication may include protocols, but not limiting to, such as bluetooth, wifi, gsm, gprs). The sub-blocks shown in Fig 3 aircraft system are not all exclusive and may increase depending on the complexity and functionality desired from the aircraft system.

### Delivery Process

Fig. 4 shows the process flow for an article delivery using the proposed aircraft and dispenser combination. Initially articles are manually or automatically loaded in the dispenser compartments and their addresses are electronically recorded and stored in the dispenser and/or aircraft memory.

In one advanced variant of the dispenser, each individual compartment can be equipped with a sensor to automatically read the pickup and/or drop address of the articles placed inside it. For example, an RFID tag system may be used for such functionality. After reading of addresses in individual compartments they are stored in the memory of the dispenser and/or the memory of the aircraft.

A software or electronic hardware will use an intelligent algorithm to compute an optimized flight plan for the aircraft covering all article pickup and(or) delivery addresses for a given dispenser load(s). The intelligent algorithm will also determine the altitudes at which the aircraft will fly in the course of its entire flight-plan.

After flight-plan generation, the dispenser is attached to the aircraft module and is ready for flight. Alternatively, the optimized flight plan may be transmitted to the dispenser and/or aircraft during the course of a flight.

The aircraft flies carrying the dispenser to the respective delivery addresses. Aircraft flight control may be remotely piloted, autonomous or semi-autonomous. While en route to any delivery location, the dispenser will send a notification to the base station and recipient of the article, informing them that it is on its way and the estimated time of arrival. In another variant, the internal algorithm of the system will have it compute the estimated delivery times for all the addresses on a given flight plan in advance, and will send the notification to the base station and recipients well in advance via mechanisms including (but not limited to) cellphone applications, sms, email etc. In a further enhanced version, recipients may choose to receive the articles at another time of the day, and the aircraft system may re-arrange its flight plan and send them a revised estimated time of arrival which is closer to, or exactly the same as their requirement.

Once the aircraft reaches the GPS coordinates corresponding to any of the drop address, its on-board flight control system will issue an alert to the base station and to the dispenser processor. The dispenser sensors will then verify the GPS coordinates of the air-borne location. If the location coordinates don't match with a pre-stored drop point address it will issue an address re-verify instruction to the aircraft processor. If the address location is a match, the sensors in the dispenser will evaluate or calculate the current altitude of the dispenser with respect to ground level below it. As required, the dispenser processor will issue an altitude correction instruction to the aircraft processor, to lower or raise the aircraft altitude and bring it in a pre-determined safe-drop altitude range. As required, the aircraft processor will control the motors and propellers to take the aircraft and dispenser in the safe-drop altitude range. Once the aircraft and dispenser reach and hover in the safe-drop altitude range, the dispenser processor will issue a delivery alert to the base station receiver and/or the aircraft and/or any registered addressee(s) for the delivery address through its communication blocks (the alert to addressee may be issued using methods such as, but not limited to, text, sms, voip, call, email). After issuing the alert the processor will activate the load/dispense block to open the orifice (wherever applicable) corresponding to the article to be delivered at that specific address. The article will either fall freely under gravity at the location or be lowered by an appropriate lowering mechanism (including but not limiting magnetic, mechanical, electronic, or any combination of the aforementioned).

For multiple consecutive delivery addresses next to each other, the intelligent flight plan algorithm will determine the most optimal altitude changes for the aircraft to save time and battery/fuel. The algorithm will decide when to fly the aircraft at a high cruising altitude and when to fly it at a safe-drop altitude.

For example, in one variant if two or more consecutive drop addresses are within a specific range, the aircraft will stay in the safe-drop altitude and navigate from one address to other till it completes all drops in the region without rising back to the cruising altitude.

In another variant, before opening the orifice or releasing an article the dispenser may seek a delivery permission from the base-station or the aircraft or any registered user. In the event of no response or a negative response, the dispenser may abort delivery an re-attempt it at a later stage.

In yet another variant the sensor payload on the aircraft and (or) the dispenser may await an approval permission from the drop address and/or the addressee. The approval process or handshake may involve, but is not limited to, matching of a specific- visual and/or acoustic, and/or thermal and/or electronic- pattern, and/or signal that the recipient drop addressee can place anywhere on its property or even a characteristic property of the address. For example, in one variant the aircraft cameras (or on-board sensors) will be used to take images (or record other characteristic signals) specific to each drop address. These images (or signals) will be saved in the flight control and base station database. After creation of the database, when a flight will arrive at the GPS coordinates of a delivery address, its sensor payload will re-take an image or capture other signals at that location. It will then compare the re-captured image (or signal) with the one stored in its database for the corresponding address. If the two match with a certain degree of confidence, the delivery mechanism will be initiated. If the match is below a certain degree of confidence, the aircraft flight control will issue an alert notification to the base station requesting for further course of action. A human or computer at the base station will take decision in such a case for the delivery to be executed or not.

The proposed system will have a provision for delivery success acknowledgement. The acknowledgement (in form of, but not limited to, sms, text, email notification or alert) will be sent to the base station (or a registered sender of the article) from the communication block of the dispenser. In an enhanced version, variant the acknowledgement alert will be issued after a receipt of confirmation from the addressee either through a smartphone application or a specific sensor handshake with an on-board sensor of the dispensor or aircraft carrying the dispensor. The addressee will also have the option of taking a bar code scan or RfiD scan of the delivered article for generating a delivery success acknowledgement. Post such a scan the acknowledgement will be sent from the scanner device of the addressee to the dispenser communication block and/or to the base station, and finally relayed to the sender of the article. In another variant, which addresses the need for delivery of sensitive documents (which must be delivered only to the intended recipient), the system may send a barcode to the recipient in advance via email, sms, or any other method. Upon approaching the delivery location, the dispenser, equipped with a barcode scanner, will scan the barcode from the recipient's phone or tablet or other media, and only then dispense the required article to the recipient.

In yet another variant, in the event of any unusual circumstance, wherein the sender or recipient requires some manual intervention or assistance, he or she may make contact with a call center at the base station (via the cellphone app, or by indicating to the system that he or she needs special assistance through a variety of ways including but not limited to, saying a pre-determined keyword). In such an instance, an operator at the base station, who can see the sender or recipient via the camera installed on the system, can also speak to them to resolve any problems they may have.

Once the article has been delivered, the orifice will be automatically closed (wherever applicable). The aircraft and dispenser will then proceed to repeat the above mentioned delivery process at the next deliver address in the aircraft's flight plan.

The aircraft unit will have intelligent awareness of its remaining battery life at any given point in time. Whenever the remaining battery life is approaching a pre-set critical 'remaining percentage of battery', the aircraft unit will send a message to the communication block of the dispenser informing it to pause all deliveries and will proceed to fly back to the base station for re-charging or retirement for the day.

### *Customization of Articles*

<u>In one variant, an additional material may be added to one or more articles which are due for delivery. This may include, but not be limited to, a device with low mass but high density compared to air. Examples include, a piece of cork, or cellophane paper, amongst others. This may enhance the positional accuracy of the article being delivered and reduce the air drag during drop.</u>

### *Crowd Sourcing*

The abovementioned system may also be used for a crowd sourcing platform wherein via a cellphone application or other system, users in need of articles to be delivered may 'hail' a dispenser to carry out the activity. The application will show the user, the nearest dispenser location and the estimated time of arrival of the dispenser at the desired location. The user may book the dispenser for either collecting articles for delivery to another location, or collecting articles from another location to bring to the user. The user will input the pick-up and drop addresses into the cellphone application, for this purpose. Upon reaching a designated pick-up location, the dispenser will land at the location. With the sensor block being aware of the available/empty compartments, one of the available/empty compartments will open from the top (top-loading), and the user can place his/her article in that compartment, after which the compartment will shut. Once the user confirms completion of the loading by pressing a designated button on the cellphone app, the dispenser will fly to the drop off location and complete the delivery as described in the earlier sections of this report. The drop off location may be a manually entered address, or a be based on the cellphone location of an intended recipient (in which case, it may change from time to time as the recipient moves). The cellphone application may also be integrated into already existing chat applications for users' convenience.

## Figures

Fig 1 – Dispenser 3D view, top and bottom faces of the dispenser and system block diagram

Fig 2 – Dispenser attached with aircraft module

Fig 3 – Interaction between aircraft system and the dispenser system

Fig 4 – Intelligent aerial article dispense methodology

Fig 5 – Diagram showing airborne aircraft and dispenser, base-station and drop location

## References & Priori Art:

## Terms

**Delivery, delivering** - releasing/dropping/lowering/unloading/loading/dispensing
**Target** - Aerial, Land or Sea based address, location, recipient

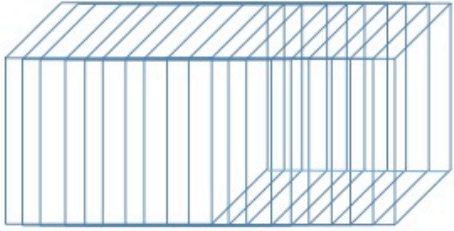

Fig. 1 a

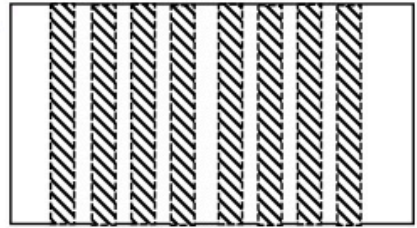

Fig. 1 b

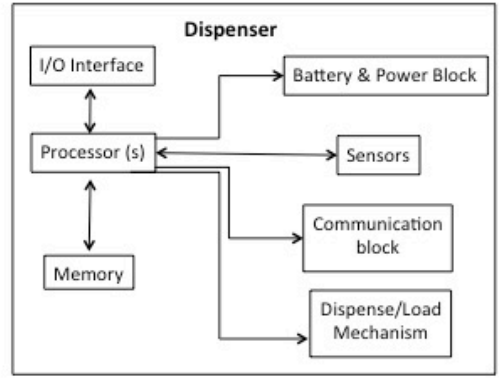

Fig. 1 c

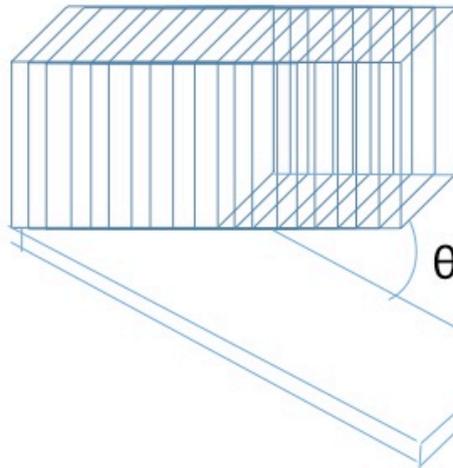

Fig. 1 d

Can be interfaced with a mailbox or some other landing point

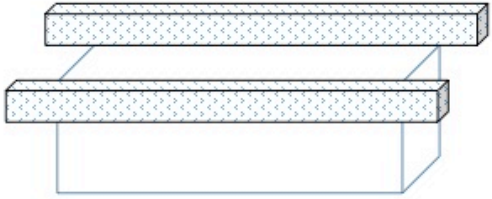

Fig. 2a

Flight system connected in partial configuration

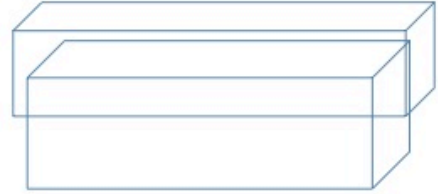

Fig. 2b

Flight system connected in all-around configuration

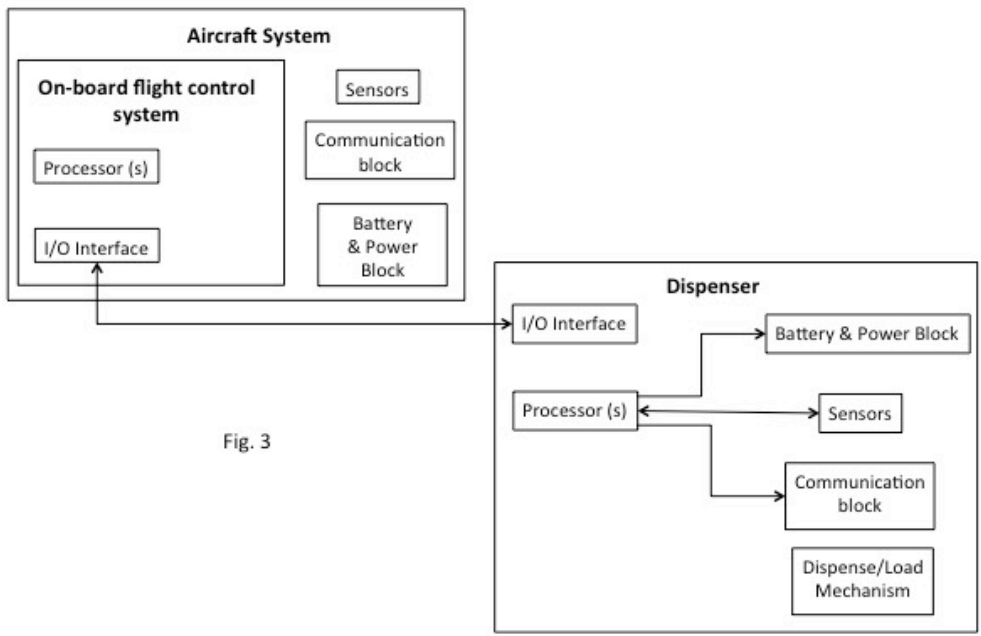

Fig. 3

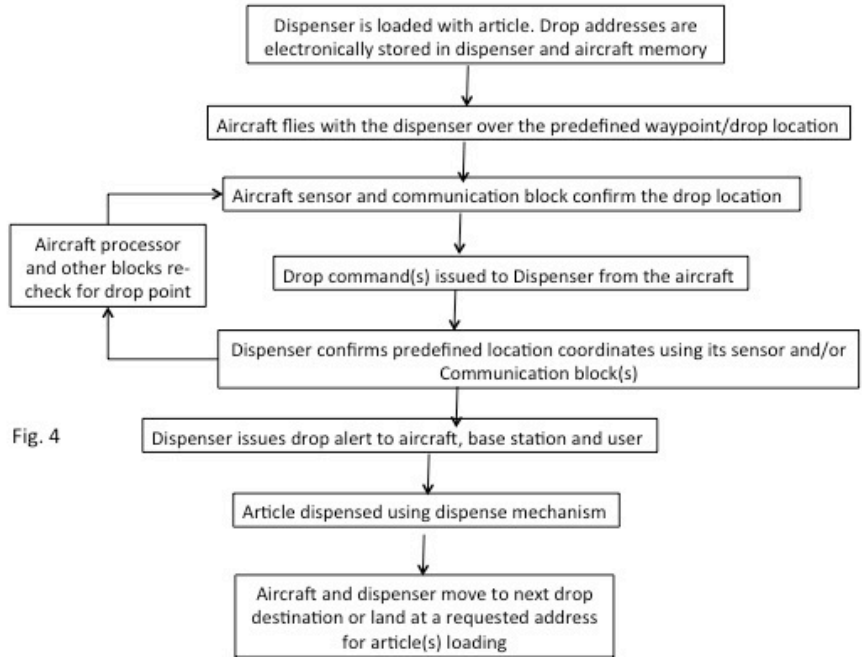

Fig. 4

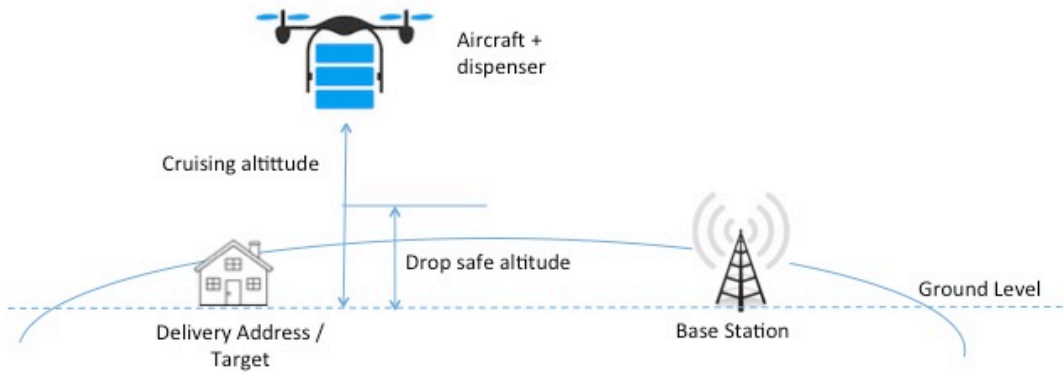

Fig. 5